\documentclass[prd,preprint,superscriptaddress,showpacs,byrevtex]{revtex4}
\usepackage{bm}
\def\fsl#1{\setbox0=\hbox{$#1$}           
   \dimen0=\wd0                                 
   \setbox1=\hbox{/} \dimen1=\wd1               
   \ifdim\dimen0>\dimen1                        
      \rlap{\hbox to \dimen0{\hfil/\hfil}}      
      #1                                        
   \else                                        
      \rlap{\hbox to \dimen1{\hfil$#1$\hfil}}   
      /                                         
   \fi}                                         %
\newcommand{\be}{\begin{equation}}
\newcommand{\ee}{\end{equation}}
\newcommand{\bea}{\begin{eqnarray}}
\newcommand{\eea}{\end{eqnarray}}
\newcommand{\beq}{\begin{equation}}
\newcommand{\eeq}{\end{equation}}
\newcommand{\beqs}{\begin{eqnarray}}
\newcommand{\eeqs}{\end{eqnarray}}

\begin{document}
\title{ Correct Definition of the Gluon Fragmentation Function at High Energy Colliders }
\author{Gouranga C Nayak }\email{nayak@max2.physics.sunysb.edu}
\affiliation{ 22 West Fourth Street \#1, Lewistown, Pennsylvania 17044, USA }
\date{\today}
\begin{abstract}
Since the definition of the gluon to hadrons fragmentation amplitude $<H,X|g>$ involves
gluon field $Q^{\mu a}(x)$ corresponding to single gluon incoming state $|g>$,
the present definition of the gluon fragmentation function at
high energy colliders [which involves non-abelian field tensor
$F^{\mu \nu a}(x)=\partial^\mu Q^{\nu a}(x) -\partial^\nu Q^{\mu a}(x)+gf^{abc}Q^{\mu b}(x)Q^{\nu c}(x)$ instead of
$Q^{\mu a}(x)$ in the initial state] is not consistent with the single gluon incoming state $|g>$.
In this paper we derive the correct definition of the gluon fragmentation function
at high energy colliders from first principles which is obtained from the single gluon incoming state $|g>$ and is gauge
invariant and is consistent with the factorization theorem in QCD.
\end{abstract}
\maketitle
\pagestyle{plain}
\pagenumbering{arabic}
\section{Introduction}
In order to study hadron production at high energy colliders one needs to know how a hadron is formed from partons.
However, this is an unsolved problem in QCD so far because the formation of hadron from partons involves
non-perturbative QCD which is not solved yet. Hence, at present, we depend on experimental data to study hadronization
in QCD.

The usual procedure in the high energy phenomenology is to determine the parton to hadron fragmentation function
from a set of experiments at some momentum transfer scale and then determine its evolution to other
momentum transfer scale by using DGLAP evolution equations \cite{dglap}. Once the fragmentation function
is determined in this way then one uses it to predict physical observable at other collider experiments.
Hence in order for this prescription to work it is necessary that these fragmentation functions
are universal, {\it i. e.}, they do not change from one collider experiment to another
collider experiment. From this point of view it is necessary to use the correct definition of the fragmentation function at
high energy colliders. If one does not use the correct definition of the fragmentation
function at high energy colliders then one will obtain divergent cross section for physical
observable at high energy colliders. This can be seen as follows.

Consider, for example, the hadron production at high energy colliders. If the factorization theorem holds
\cite{sterman,collins,cs2,bodwin} then the formula for the hadron production cross section at high energy hadronic colliders is given by
\bea
\sigma (AB \rightarrow H+X) ~=~ \sum_{a,b}~\int dx_1 ~\int dx_2 ~\int dz~f_{a/A}(x_1,Q^2)~f_{b/B}(x_2,Q^2)~{\hat \sigma}(ab \rightarrow cd)~D_{H/c}(z,Q^2) \nonumber \\
\label{cr1v}
\eea
where $f_{a/A}(x_{1},Q^2)$ (or $f_{b/B}(x_{2},Q^2)$)
is the parton distribution function (PDF), ${\hat \sigma}(ab \rightarrow cd)$ is the partonic level scattering
cross section and $D_{H/c}(z,Q^2)$ is the fragmentation function (FF) where $a,b,c,d=q,{\bar q},g$ represent quark, antiquark, gluon.

Hence if one does not use the correct definition of the fragmentation function
in eq. (\ref{cr1v}) then any uncanceled infrared divergences in the Feynman diagrams in the
partonic level scattering cross section ${\hat \sigma}(ab \rightarrow cd)$ can not be absorbed in the definition of the non-perturbative
fragmentation function $D_{H/a}(z,Q^2)$ in eq. (\ref{cr1v}). Hence it is necessary to
derive the correct definition of the fragmentation function $D_{H/a}(z,Q^2)$ from first principles.

A correct definition of a parton to hadron fragmentation function at high energy colliders must satisfy the following three properties:
1) since we are considering a single parton fragmenting to hadrons, it
must be obtained from the single parton state $|c>$, 2) it must be gauge invariant and 3)
it must be consistent with factorization theorem in QCD. The present definition of the quark to hadron fragmentation function
at high energy colliders is consistent with these three properties \cite{collins}. However, the present definition of the gluon
to hadron fragmentation function at high energy colliders \cite{collins} is not consistent with these three properties as we will
see below.

The present definition of the gluon to hadron fragmentation function at high energy colliders is given by \cite{collins}
\bea
&& D_{H/g}(z)
= \frac{z}{32\pi k^+} ~\int dy^- e^{-i{P}^+ y^-/z} \nonumber \\
&&[<0| F^{+\mu a}(0) ~\times ~ {\cal P}{\rm exp}[-igT^{(A)b}\int_0^{\infty}dz^-Q^{+b}(0,z^-,0_T)]~a^\dagger_H(P^+,0_T)  a_H(P^+,0_T)\nonumber \\
&&~\times ~ {\cal P}{\rm exp}[-igT^{(A)c}\int_{0}^{\infty}dz^-Q^{+c}(0,y^-+z^-,0_T)]\times ~F_\mu^{~+a}(0,y^-,0_T)  |0>]
\label{gpdfv}
\eea
where
\bea
F_{\mu \nu}^a(x)=\partial_\mu Q_\nu^a(x)-\partial_\nu Q_\mu^a(x)+gf^{abc}Q_\mu^b(x)Q_\nu^c(x),
\label{ffmnxv}
\eea
$T^{(A)a}_{bc}=-if^{abc}$ and $Q^{\mu a}(x)$ is the quantum gluon field.

The expression ${\cal P}{\rm exp}[igT^{(A)b}\int_{z_1^-}^{z_2^-} dz^-Q^{+b}(0,z^-,0_T)]$ in eq. (\ref{gpdfv})
stands for the light-like Wilson line. Since the light-like Wilson line in QCD involves
soft-collinear gluon field $A^{\mu a}(x)$, it can be treated classically as $A^{\mu a}(x)$ becomes
the SU(3) pure gauge when Wilson line becomes light-like (see section III for details).
Hence for the better understanding of the single gluon initial state $|g>$
in eq. (\ref{gpdfv}) let us replace the light-like Wilson line
${\cal P}{\rm exp}[igT^{(A)b}\int_{z_1^-}^{z_2^-} dz^-Q^{+b}(0,z^-,0_T)]$ in eq. (\ref{gpdfv})
by its classical counterpart ${\cal P}{\rm exp}[igT^{(A)b}\int_{z_1^-}^{z_2^-} dz^-A^{+b}(0,z^-,0_T)]$
where $A^{\mu a}(x)$ is the classical SU(3) pure gauge background field.
With this one finds from eq. (\ref{ffmnxv}) that $F^{\mu \nu a}(x)F_{\mu \nu}^a(0)$
in eq. (\ref{gpdfv}) contains quadratic, cubic and quartic powers of the (quantum) gluon field $Q^{\mu a}(x)$ which implies that the
present definition of gluon fragmentation function at high energy colliders
in eq. (\ref{gpdfv}) is not consistent with the single gluon initial state $|g>$,
{\it i.e.}, $F^{\mu \nu a}(x)F_{\mu \nu}^a(0)\not\propto Q^{\mu a}(x)Q_\mu^a(0)$ in eq. (\ref{gpdfv}) in QCD.
This implies that the present definition of the gluon fragmentation function at high energy colliders is not correct.

In this paper we derive correct definition of the gluon fragmentation function at high energy colliders from first
principles which is obtained from the single gluon initial state $|g>$ and is gauge invariant and is
consistent with the factorization theorem in QCD.

We find that the correct definition of the gluon fragmentation function at high energy colliders
obtained from the single gluon initial state $|g>$ and is gauge invariant and is consistent with the
factorization theorem in QCD is given by
\bea
&&D_{H/g}(z)
= \frac{zk^+}{32\pi } ~\int dx^- e^{i{P}^+ x^-/z} \nonumber \\
&&[<0| Q^{\mu a}(x^-) ~\times ~ {\cal P}{\rm exp}[-igT^{(A)b}\int_0^{\infty}dz^-A^{+b}(x^-+z^-)]~a^\dagger_H(P^+,0_T)  a_H(P^+,0_T)\nonumber \\
&&~\times ~ {\cal P}{\rm exp}[-igT^{(A)c}\int_{0}^{\infty}dz^-A^{+c}(z^-)]\times ~Q_\mu^{a}(0)  |0>]
\label{gpdffpiv}
\eea
which is valid in covariant gauge, in light-cone gauge, in general axial gauges, in general non-covariant gauges and in
general Coulomb gauge etc. respectively. Note that the correct definition of the gluon fragmentation function
at high energy colliders in eq. (\ref{gpdffpiv}) is gauge invariant with respect to the gauge transformation
\bea
T^aA'^a_\mu(x) = U(x)T^aA^a_\mu(x) U^{-1}(x)+\frac{1}{ig}[\partial_\mu U(x)] U^{-1}(x),~~~~~~~~~~~U(x)=e^{igT^a\omega^a(x)}
\label{aftgrmpiv}
\eea
along with the homogeneous transformation \cite{thooft,zuber,zuber1,abbott}
\bea
T^aQ'^a_\mu(x)=U(x)T^aQ^a_\mu(x) U^{-1}(x),~~~~~~~~~~~U(x)=e^{igT^a\omega^a(x)}
\label{jprtpiv}
\eea
where $Q^{\mu a}(x)$ is the hard (quantum) gluon field which fragments to hadron $H$
and $A^{\mu a}(x)$ is the soft-collinear gluon field which is the SU(3) pure gauge background field.
After renormalization the gluon fragmentation function
is expected to obey a QCD evolution equation, like DGLAP equation \cite{tung}, which follows from renormalization group
equation.

We will provide a derivation of eq. (\ref{gpdffpiv}) in this paper.

The paper is organized as follows. In section II we discuss the field theoretical derivation
of the definition of the gauge non-invariant gluon fragmentation function in QCD.
In section III we discuss the gauge invariant non-perturbative gluon correlation function and proof of
factorization theorem in QCD. In section IV we derive the correct definition of the gluon fragmentation
function at high energy colliders as given by eq. (\ref{gpdffpiv}). Section V contains conclusions.

\section{ Definition of the Gauge Non-Invariant Gluon Fragmentation Function in QCD }

In this section we will present the  derivation of the definition of the
gauge non-invariant gluon to hadron fragmentation function in QCD. The derivation
of the definition of the
gauge invariant gluon to hadron fragmentation function at high energy colliders
which is consistent with factorization theorem in QCD will be discussed in next
sections. For application to collider experiments it is convenient to use
light cone quantization formalism. To introduce the concept of derivation of the
definition of the fragmentation function using quantum field theory
we will consider the scalar gluon first before considering the
gluon in QCD. Note that the scalar gluon is not physical. The only reason we have considered it here
is for simplicity in order to motivate the similar derivation of the definition of the gauge non-invariant
gluon to hadron fragmentation function in QCD.

\subsection{Definition of Scalar Gluon Fragmentation Function Using Quantum Field Theory }

Consider a scalar gluon with four momentum $k^\mu$ in vacuum fragmenting to a hadron with four
momentum $P^\mu$. Note that the derivation of the definition of the scalar gluon fragmentation
function is described in \cite{collins,nayakfg} which we will briefly review here in order to
motivate the similar derivation of the definition of the gauge non-invariant
gluon to hadron fragmentation function in QCD. In the light-cone coordinate system the
scalar gluon field $\phi(x^-,x_T)$ can be written as
\bea
\phi(x^-,x_T) = \frac{1}{(2\pi)^{3}} ~\int \frac{dk^+}{\sqrt{2k^+}} d^{2}k_T~[e^{-ik^+x^-+ik_T \cdot x_T} a(k^+,k_T) + e^{ik^+x^--ik_T \cdot x_T} a^\dagger (k^+,k_T)] \nonumber \\
\label{phii}
\eea
where $a$ is the annihilation operator and $a^\dagger$ is the creation operator.
The single particle scalar gluon initial state $|k^+, k_T>$ can be obtained from the vacuum state $|0>$
by using
\bea
|k^+, k_T> = a^\dagger (k^+, k_T) |0>
\label{kveci}
\eea
where
\bea
a(k^+, k_T) |0> =0.
\label{kvecbi}
\eea
The normalization condition is given by
\bea
<k^+, k_T|{k'}^+, {k'}_T> = (2\pi)^{3} \delta(k^+-{k'}^+) \delta^{(2)}(k_T -{k'}_T).
\label{normi}
\eea
As explained in \cite{nayakfg} the correct interpretation of the state $|k^+,k_T>$ is created by an appropriate
Fourier transform of the corresponding field operator and should not be associated
with on-shell condition $k^2=0$ of the gluon.

For the inclusive production of a hadron $H$ created in the $out-$state $|H,X>$ from
the scalar gluon in the $in-$state $|k^+,k_T>$ the probability amplitude is given by
\bea
<H,X|k^+,k_T>
\eea
which gives the probability distribution $h_k(P)$ of the hadron with momentum $P$ from the parton of
momentum $k$
\bea
&& h_k(P) <k^+,k_T|k'^+,k'_T>=\sum_X~<k^+,k_T|H,X><H,X|k'^+,k'_T>\nonumber \\
&& =\sum_X~<k^+,k_T|a^\dagger_H(P)|X><X|a_H(P)|k'^+,k'_T> =<k^+,k_T|a^\dagger_H(P)a_H(P)|k'^+,k'_T>\nonumber \\
\label{hp2i}
\eea
where $a^\dagger_H$ is the creation operator of the hadron $H$. Hence we find
(by using eqs. (\ref{kveci}) and (\ref{normi})) that
\bea
&& h(z,P_T) <k^+,k_T|{k'}^+,{k'}_T>=2z(2\pi)^{3}D_{H/a}(z,P_T) (2\pi)^{3} \delta(k^+-{k'}^+) \delta^{(2)}(k_T -{k'}_T) \nonumber \\
&& =<0|a(k^+,k_T)a^\dagger_H(P^+,P_T)a_H(P^+,P_T)a^\dagger({k'}^+,{k'}_T)|0>
\label{jetfi}
\eea
where $D_{H/a}(z,P_T)$ is the transverse momentum dependent fragmentation function and $z=\frac{P^+}{k^+}$
is the longitudinal momentum fraction of hadron $H$ with respect to parton $a$.
Using
\bea
&& (2\pi)^{3} <0| \phi(x^-,x_T) a^\dagger_H(P^+,P_T)  a_H(P^+,P_T) \phi(0) |0>
 = \frac{1}{(2\pi)^{3}} \int \frac{dk^+}{\sqrt{2k^+}} d^{2}k_T \int \frac{d{k'}^+}{\sqrt{2{k'}^+}} d^{2}{k'}_T \nonumber \\
 &&~<0| e^{-ik^+x^-+ik_T\cdot x_T} a(k^+,k_T)a^\dagger_H(P^+,P_T)a_H(P^+,P_T)a^\dagger({k'}^+,{k'}_T)|0>
\label{fr3i}
\eea
we find from eq. (\ref{jetfi})
\bea
D_{H/a}(z,P_T)
= \frac{k^+}{z} \int dx^- \frac{d^{2}x_T}{(2\pi)^{3}}  e^{i{k}^+ x^- - i {k}_T \cdot x_T} <0| \phi(x^-,x_T) a^\dagger_H(P^+,P_T)  a_H(P^+,P_T) \phi(0) |0>.~~~~~
\label{fr6oli}
\eea
We rewrite the definition in a form analogous to the definition of the distribution of
partons in a hadron where the transverse momentum is of the parton relative to the hadron rather than vice versa.
Hence making a Lorentz transformation to a frame where the hadron's transverse momentum is
zero:
\bea
&& (P^+, P_T) \equiv (P^+,0)  \nonumber \\
&& (k^+,0) \equiv (k^+, -P_T/z).
\label{mti}
\eea
we find from eq. (\ref{fr6oli})
\bea
D_{H/a}(z,P_T)
= \frac{k^+}{z} \int dx^- \frac{d^{2}x_T}{(2\pi)^{3}}  e^{i{k}^+ x^- + i {P}_T \cdot x_T/z}
<0| \phi(x^-,x_T) a^\dagger_H(P^+,0_T)  a_H(P^+,0_T) \phi(0) |0>.~~~~~~~
\label{fr6i}
\eea
Integrating over $p_T$ we find that the scalar gluon to hadron fragmentation function is given by
\bea
&& D_{H/a} (z)=\frac{zk^+}{2\pi}~\int dx^-e^{i{P}^+x^-/z}
<0|\phi(x^-) a^\dagger_H(P^+,0_T)a_H(P^+,0_T) \phi(0) |0>.
\label{frz2i}
\eea

\subsection{ Definition of Gauge Non-Invariant Gluon Fragmentation Function in QCD }

In analogous to eq. (\ref{frz2i}) for the scalar gluon fragmentation function
we find that the definition of the gauge non-invariant gluon
to hadron fragmentation function in QCD is given by
\bea
&&D_{H/g}(z)
= \frac{zk^+}{32\pi } ~\int dx^- e^{i{P}^+ x^-/z} <0| Q^{\mu a}(x^-)a^\dagger_H(P^+,0_T)  a_H(P^+,0_T) Q_\mu^{a}(0)  |0>
\label{gfrg}
\eea
which contains $Q^{\mu a}(x) Q_\mu^{a}(0)$ which is proportional to the quadratic power of the (quantum) gluon field
$Q^{\mu a}(x)$ instead of the non-abelian field tensor $F^{\mu \nu a}(x)F_{\mu \nu}^a(0)$ as in eq. (\ref{gpdfv})
which contains quadratic, cubic and quartic powers of the (quantum) gluon field where
$F_{\mu \nu}^a(x)$ is given by eq. (\ref{ffmnxv}). Hence we find that, unlike eq. (\ref{gpdfv}),
the gauge non-invariant definition of the gluon to hadron fragmentation function in eq. (\ref{gfrg}) is obtained
from the single gluon initial state $|g>$ in QCD.

In order to obtain gauge invariant definition of the gluon
to hadron fragmentation function which is consistent with
factorization theorem in QCD we need to incorporate Wilson lines appropriately
in the definition of gauge non-invariant gluon to hadron fragmentation function in QCD in eq. (\ref{gfrg}).
Hence we need to prove factorization theorem of the non-perturbative gluon correlation function in QCD.

\section{ Gauge Invariant Non-perturbative gluon correlation function and proof of factorization theorem in QCD }

As mentioned above the definition of the gluon fragmentation function in eq. (\ref{gfrg}) is not gauge invariant.
Hence it can not cancel any uncanceled infrared divergences in the perturbative cross section ${\hat \sigma}(ab \rightarrow cd)$
in eq.  (\ref{cr1v}). This implies that we need to prove factorization theorem for the infrared divergences
of the non-perturbative gluon correlation
function of the type $<0|Q_\mu^a(x_1)Q_\nu^b(x_2)|0>$ in QCD in order to obtain the gauge invariant definition
of the gluon fragmentation function from eq. (\ref{gfrg}).

The infrared divergences can be studied by using eikonal approximation in quantum field theory.
The Eikonal propagator times the Eikonal vertex for a gluon with four momentum $k^\mu$ interacting with a
light-like quark moving with four velocity $l^\mu$ is given by
\cite{collins,tucci,collinssterman,berger,frederix,nayakqed,scet1,pathorder,nayaka2,nayaka3,nayaksterman,nayaksterman1}
\bea
gT^a~\frac{l^\mu}{l \cdot k+i\epsilon }.
\label{eikonalinp}
\eea
Hence for a light-like quark attached to infinite number of gluons we find the eikonal factor
\bea
&& 1+ gT^a\int \frac{d^4k}{(2\pi)^4} \frac{l\cdot { A}^a(k)}{l\cdot k +i\epsilon }
+g^2\int \frac{d^4k_1}{(2\pi)^4} \frac{d^4k_2}{(2\pi)^4} \frac{T^a l\cdot { A}^a(k_1)T^b l\cdot { A}^b(k_2)}{(l\cdot k_1 +i \epsilon)(l\cdot (k_1+k_2) +i \epsilon)}+... \nonumber \\
&&=1+ igT^a\int_0^{\infty} d\lambda l\cdot {  A}^a(l\lambda)+g^2i^2 \int_0^{\infty}  d\lambda_1 \int_{\lambda_1}^{\infty} d\lambda_2 T^a l\cdot { A}^a(l\lambda_1) T^b l\cdot { A}^b(l\lambda_2)+...\nonumber \\
&&= 1+ igT^a\int_0^{\infty} d\lambda l\cdot {  A}^a(l\lambda)+\frac{g^2i^2}{2!} {\cal P}\int_0^{\infty}  d\lambda_1 \int_0^{\infty}  d\lambda_2 T^a l\cdot { A}^a(l\lambda_1) T^b l\cdot { A}^b(l\lambda_2)+...
\nonumber \\
&&={\cal P}~{\rm exp}[ig \int_0^{\infty} d\lambda l\cdot { A}^a(l\lambda)T^a ]
\label{iiij}
\eea
which exactly describes the infrared divergences arising from the infinite number of soft gluons exchange
with the light-like quark where ${\cal P}$ is  the path ordering and the gluon field $ { A}^{\mu a}(x)$ and its Fourier transform
$ { A}^{\mu a}(k)$ are related by
\bea
{ A}^{\mu a}(x) =\int \frac{d^4k}{(2\pi)^4} { A}^{\mu a}(k) e^{ik \cdot x}.
\label{ft}
\eea.
A light-like quark traveling with light-like four-velocity $l^\mu$ produces SU(3) pure gauge potential $A^{\mu a}(x)$
at all the time-space position $x^\mu$ except at the position ${\vec x}$ perpendicular to the direction of motion
of the quark (${\vec l}\cdot {\vec x}=0$) at the time of closest approach \cite{collinssterman,nayakj,nayake}.
When $A^{\mu a}(x) = A^{\mu a}(\lambda l)$ as in eq. (\ref{iiij})
we find ${\vec l}\cdot {\vec x}=\lambda {\vec l}\cdot {\vec l}=\lambda\neq 0$ which implies that the light-like quark
finds the gluon field $A^{\mu a}(x)$ in eq. (\ref{iiij}) as the SU(3) pure gauge. The SU(3) pure gauge is given by
\bea
T^aA_\mu^a (x)= \frac{1}{ig}[\partial_\mu U(x)] ~U^{-1}(x),~~~~~~~~~~~~~U(x)=e^{igT^a\omega^a(x)}
\label{gtqcd}
\eea
which gives
\bea
U(x_f)={\cal P}e^{ig \int_{x_i}^{x_f} dx^\mu A_\mu^a(x) T^a}U(x_i)=e^{igT^a\omega^a(x_f)}.
\label{uxf}
\eea
Hence when $A^{\mu a}(x) = A^{\mu a}(\lambda l)$ as in eq. (\ref{iiij}) we find from eq.
(\ref{uxf}) that the light-like Wilson line in QCD for infrared divergences is given by
\cite{nayaknrqcd,nayakg,nayakg1,nayakg2}
\bea
\Phi(x)={\cal P}e^{-ig \int_0^{\infty} d\lambda l\cdot { A}^a(x+l\lambda)T^a }=e^{igT^a\omega^a(x)}.
\label{ttt}
\eea
In the adjoint representation of SU(3) the corresponding path ordered exponential is given by
\bea
\Phi_{ab}(x)={\cal P}{\rm exp}[-ig\int_0^{\infty} d\lambda l\cdot { A}^c(x+l\lambda)T^{(A)c}],~~~~~~~~~(T^{(A)c})_{ab}=-if^{abc}
\label{wilabfv}
\eea
which under non-abelian gauge transformation, as given by eq. (\ref{aftgrmpiv}), transforms as
\bea
\Phi'_{ab}(x)=U_{adj}(x){\cal P}{\rm exp}[-ig\int_0^{\infty} d\lambda l\cdot { A}^c(x+l\lambda)T^{(A)c}],~~~~~~~~~U_{adj}(x)=e^{igT^{(A)b}\omega^b(x)}.
\label{ty}
\eea
Hence the effect of infrared gluons interaction between the partons and the light-like Wilson line in QCD can be studied by
putting the partons in the SU(3) pure gauge background field which implies that the infrared behavior of the non-perturbative
gluon correlation function of the type $<0|Q_\mu^a(x) Q_\nu^b(x') Q_\lambda^c(x'')...|0>$ in QCD
due to the presence of light-like Wilson line in QCD can be studied by using the path integral method
of the QCD in the presence of SU(3) pure gauge background field as given by eq. (\ref{gtqcd}).

The non-perturbative gluon correlation function of the type $<0|Q_\mu^a(x_1) Q_\nu^b(x_2)|0>$ in QCD is given by \cite{tucci,muta,abbott}
\bea
&&<0|Q_\mu^a(x_1) Q_\nu^b(x_2)|0>=\int [dQ] [d{\bar \psi}] [d \psi ] ~Q_\mu^a(x_1) Q_\nu^b(x_2)\nonumber \\
&& \times {\rm det}(\frac{\delta (\partial_\mu Q^{\mu a})}{\delta \omega^b})~
e^{i\int d^4x [-\frac{1}{4}{F^a}_{\mu \nu}^2[Q] -\frac{1}{2 \alpha}(\partial_\mu Q^{\mu a})^2+{\bar \psi} [i\gamma^\mu \partial_\mu -m +gT^a\gamma^\mu Q^a_\mu] \psi  ]}
\label{cfq5}
\eea
where the suppression of the normalization factor $Z[0]$ is understood as it will cancel in the final result
[see eq. (\ref{finalzv})].

Background field method of QCD was originally formulated by 't Hooft \cite{thooft} and later
extended by Klueberg-Stern and Zuber \cite{zuber,zuber1} and by Abbott \cite{abbott}.
This is an elegant formalism which can be useful to construct gauge invariant
non-perturbative green's functions in QCD. This formalism is also useful to study quark and gluon production from classical chromo field \cite{peter}
via Schwinger mechanism \cite{schw}, to compute $\beta$ function in QCD \cite{peskin}, to perform
calculations in lattice gauge theories \cite{lattice} and to study evolution of QCD
coupling constant in the presence of chromofield \cite{nayak}.

The non-perturbative gluon correlation function of the type $<0|Q_\mu^a(x_1) Q_\nu^b(x_2)|0>_A$
in the background field method of QCD is given by \cite{tucci,thooft,abbott,zuber,zuber1}
\bea
&&<0|Q_\mu^a(x_1) Q_\nu^b(x_2)|0>_A=\int [dQ] [d{\bar \psi}] [d \psi ] ~Q_\mu^a(x_1) Q_\nu^b(x_2)\nonumber \\
&& \times {\rm det}(\frac{\delta G^a(Q)}{\delta \omega^b}) e^{i\int d^4x [-\frac{1}{4}{F^a}_{\mu \nu}^2[A+Q] -\frac{1}{2 \alpha}
(G^a(Q))^2+{\bar \psi} [i\gamma^\mu \partial_\mu -m +gT^a\gamma^\mu (A+Q)^a_\mu] \psi   ]}
\label{cfqcd}
\eea
where the suppression of the normalization factor $Z[0]$ is understood as it will cancel in the final result
[see eq. (\ref{finalzv})]. In eq. (\ref{cfqcd}) the gauge fixing term is given by
\bea
G^a(Q) =\partial_\mu Q^{\mu a} + gf^{abc} A_\mu^b Q^{\mu c}=D_\mu[A]Q^{\mu a}
\label{gav}
\eea
which depends on the background field $A^{\mu a}(x)$ and
\bea
F_{\mu \nu}^a[A+Q]=\partial_\mu [A_\nu^a+Q_\nu^a]-\partial_\nu [A_\mu^a+Q_\mu^a]+gf^{abc} [A_\mu^b+Q_\mu^b][A_\nu^c+Q_\nu^c].
\eea

When the background field is the SU(3) pure gauge as given by eq. (\ref{gtqcd}) then we find that \cite{nayaknrqcd,nayakg,nayakg1,nayakg2}
\bea
<0|Q_\mu^a(x_1)Q_\nu^b(x_2)|0> =<0|\Phi_{ac}(x_1)Q_\mu^c(x_1) \Phi_{bd}(x_2)Q_\nu^d(x_2)|0>_A
\label{finalzv}
\eea
which proves factorization of infrared divergences at all order in coupling constant in QCD
where $\Phi_{ab}(x)$ is the non-abelian phase factor or eikonal factor in the adjoint
representation of SU(3) as given by eq. (\ref{wilabfv}). Note that
eq. (\ref{finalzv}) is valid in covariant gauge, in light-cone gauge, in general axial gauges, in general
non-covariant gauges and in general Coulomb gauge etc. respectively \cite{nayakg,nayakg2}.
From eqs. (\ref{jprtpiv}) and (\ref{ty}) we find that
$<0|\Phi_{ac}(x_1)Q_\mu^c(x_1) \Phi_{bd}(x_2)Q_\nu^d(x_2)|0>_A$ in eq. (\ref{finalzv})
is gauge invariant and eq. (\ref{finalzv}) is consistent with the factorization of infrared
divergences at all order in coupling constant in QCD.

\section{ Correct Definition of the Gluon Fragmentation Function at High Energy Colliders }

From eqs. (\ref{gfrg}), (\ref{finalzv}) and (\ref{wilabfv}) we find that the correct
definition of the gluon fragmentation function at high energy colliders
obtained from the single gluon initial state $|g>$ and is gauge invariant and is consistent with the
factorization theorem in QCD is given by
\bea
&&D_{H/g}(z)
= \frac{zk^+}{32\pi } ~\int dx^- e^{i{P}^+ x^-/z} \nonumber \\
&&[<0| Q^{\mu a}(x) {\cal P}{\rm exp}[-ig\int_0^{\infty} d\lambda l\cdot { A}^b(x+l\lambda)T^{(A)b}]a^\dagger_H(P^+,0_T)  a_H(P^+,0_T)\nonumber \\
&&{\cal P}{\rm exp}[-ig\int_0^{\infty} d\lambda l\cdot { A}^c(l\lambda)T^{(A)c}]Q_\mu^{a}(0)  |0>]_{x^+=x_T=0}
\label{fffv}
\eea
which is valid in covariant gauge, in light-cone gauge, in general axial gauges, in general
non-covariant gauges and in general Coulomb gauge etc. respectively. Eq. (\ref{fffv}) can be written as
\bea
&&D_{H/g}(z)
= \frac{zk^+}{32\pi } ~\int dx^- e^{i{P}^+ x^-/z} \nonumber \\
&&<0| Q^{\mu a}(x^-) ~\times ~ {\cal P}{\rm exp}[-igT^{(A)b}\int_0^{\infty}dz^-A^{+b}(x^-+z^-)]~a^\dagger_H(P^+,0_T)  a_H(P^+,0_T)\nonumber \\
&&~\times ~ {\cal P}{\rm exp}[-igT^{(A)c}\int_{0}^{\infty}dz^-A^{+c}(z^-)]\times ~Q_\mu^{a}(0)  |0>
\label{gpdffpfv}
\eea
which reproduces eq. (\ref{gpdffpiv}).

This completes the derivation of the correct definition of the gluon fragmentation function at high energy colliders from first principles.

\section{Conclusions}
Since the definition of the gluon to hadrons fragmentation amplitude $<H,X|g>$ involves
gluon field $Q^{\mu a}(x)$ corresponding to single gluon incoming state $|g>$,
the present definition of the gluon fragmentation function at
high energy colliders [which involves non-abelian field tensor
$F^{\mu \nu a}(x)=\partial^\mu Q^{\nu a}(x) -\partial^\nu Q^{\mu a}(x)+gf^{abc}Q^{\mu b}(x)Q^{\nu c}(x)$ instead of
$Q^{\mu a}(x)$ in the initial state] is not consistent with the single gluon incoming state $|g>$.
In this paper we have derived the correct definition of the gluon fragmentation function
at high energy colliders from first principles which is obtained from the single gluon incoming state $|g>$ and is gauge
invariant and is consistent with the factorization theorem in QCD.


\begin{thebibliography}{99}

\bibitem{dglap} V. N. Gribov and L. N. Lipatov, Sov. J. Nucl. Phys. 15 (1972) 438;
L. N. Lipatov, Sov. J. Nucl. Phys. 20 (1975) 94; G. Altarelli and G. Parisi, Nucl.
Phys. B126 (1977) 298; Yu. L. Dokshitzer, Sov. Phys. J. Exp. Theor. Phys. 46 (1977) 641.

\bibitem{sterman} J. C. Collins, D. E. Soper and G. Sterman, Nucl. Phys. B223 (1983) 381; 261 (1985)
104.

\bibitem{collins} J. C. Collins and D. E. Soper, Nucl. Phys. B193 (1981) 381; B194 (1982) 445.

\bibitem{cs2} J. Collins, D. E. Soper and G. Sterman, Phys. Lett. 109B (1982) 388;
126B (1983) 275; 134B (1984) 263.

\bibitem{bodwin} G. Bodwin, Phys. Rev. D31 (1985) 2616.

\bibitem{abbott} L. F. Abbott, Nucl. Phys. B185 (1981) 189.

\bibitem{thooft} G. 't Hooft, Nucl. Phys. B62 (1973) 444.

\bibitem{zuber} H. Klueberg-Stern and J. B. Zuber, Phys. Rev. D12 (1975) 482.

\bibitem{zuber1} H. Klueberg-Stern and J. B. Zuber, Phys. Rev. D12 (1975) 3159.

\bibitem{tung} W-K. Tung, {\it Perturbative QCD and the parton structure of the nucleon}, Based on a contribution to "At the frontier of particle physics|Handbook of QCD, vol. 2* 887-971, ed. M. Shifman, World Scientific, 2001".

\bibitem{nayakfg} G. C Nayak, Eur. Phys. J. C59(2009)891, arXiv:0808.1288 [hep-ph].

\bibitem{collinssterman} J. C. Collins, D. E. Soper and G. Sterman, Nucl. Phys. B261 (1985) 104.

\bibitem{tucci} R. Tucci, Phys. Rev. D32 (1985) 945.

\bibitem{nayakqed} G. C. Nayak, Annals Phys. 324 (2009) 2579.

\bibitem{scet1} C. W. Bauer, D. Pirjol and I. W. Stewart, Phys.Rev.D65 (2002) 054022, hep-ph/0109045.

\bibitem{pathorder} J. Chay, C. Kim, Y. G. Kim, J-P. Lee, Phys.Rev. D71 (2005) 056001, hep-ph/0412110.

\bibitem{berger} C. F. Berger, hep-ph/0305076.

\bibitem{nayaka2} G. C. Nayak, Annals Phys. 325 (2010) 514.

\bibitem{nayaka3} G. C. Nayak, Annals Phys. 325 (2010) 682.

\bibitem{frederix} R. Frederix, "{\it Wilson lines in QCD}", nikhef/masters-thesis (2005).

\bibitem{nayaksterman} G. C. Nayak, J. Qiu and G. Sterman, Phys. Lett. B613 (2005) 45; Phys.Rev. D72 (2005) 114012;
Phys.Rev. D74 (2006) 074007.

\bibitem{nayaksterman1} G. C. Nayak, J. Qiu and G. Sterman, Phys.Rev.Lett. 99 (2007) 212001; Phys.Rev. D77 (2008) 034022.

\bibitem{nayakj} G. C. Nayak, JHEP03(2013)001.

\bibitem{nayake} G. C. Nayak, Eur. Phys. J. C73(2013)2442.

\bibitem{nayaknrqcd} G. C. Nayak, arXiv:1506.02591 [hep-ph].

\bibitem{nayakg} G. C. Nayak, arXiv:1506.02651 [hep-ph].

\bibitem{nayakg1} G. C. Nayak, arXiv:1506.07146 [hep-ph].

\bibitem{nayakg2} G. C. Nayak, arXiv:1508.01194 [hep-ph].

\bibitem{muta} {\it See for example}, T. Muta, {\it Foundations of Quantum Chromodynamics}, World Scientific lecture notes in physics-Vol. 5.

\bibitem{peter}  G. C. Nayak and P. van Nieuwenhuizen,
Phys. Rev. D71 (2005) 125001; G. C. Nayak, Phys. Rev. D72 (2005) 125010.

\bibitem{schw} J. Schwinger, Phys. Rev. 82 (1951) 664.

\bibitem{peskin} M. E. Peskin and D. V. Schroeder, {\it Introduction to
Quantum Field Theory}, Perseus Books Publishing, L.L.C.

\bibitem{lattice} R. F. Dashen and D. J. Gross, Phys. Rev. D23 (1981) 2340.

\bibitem{nayak} G. C. Nayak, Eur. Phys. J.C64:73,2009, arXiv:0812.5054 [hep-ph].

\end{thebibliography}
\end{document}